\documentclass[twocolumn,aps,pra,showpacs,superscriptaddress,floatfix]{revtex4}
\usepackage{graphicx}
\usepackage{times}
\usepackage{nicefrac}
\usepackage{amsmath}
\usepackage{amsfonts}
\usepackage{amssymb}
\usepackage{amsthm}
\usepackage{epsf}
\usepackage{bm}
\usepackage{bbm}

\usepackage{dcolumn}
\newcolumntype{.}{D{x}{}{-1}}
\newcommand{\qsla}{q\hspace{-0.45em} / }

\begin{document}

\title{Complete two-loop correction to the bound-electron $\bm{g}$ factor}

\author{Krzysztof Pachucki}
\affiliation{Institute of Theoretical Physics, Warsaw University,
ul.~Ho\.{z}a 69, 00--681 Warsaw, Poland}

\author{Andrzej Czarnecki}
\affiliation{Department of Physics, University of Alberta,
  Edmonton, AB, Canada T6G 2J1}

\author{Ulrich D.~Jentschura}
\affiliation{Max--Planck--Institut f\"ur Kernphysik,
Saupfercheckweg 1, 69117 Heidelberg, Germany}

\author{Vladimir A. Yerokhin}
\affiliation{Department of Physics, St.~Petersburg State University,
Oulianovskaya 1, Petrodvorets, St.~Petersburg 198504, Russia}
\affiliation{Center for Advanced Studies, St.~Petersburg State
Polytechnical University, Polytekhnicheskaya 29, 
St.~Petersburg 195251, Russia}

\begin{abstract}
Within a systematic approach based on the dimensionally regularized
nonrelativistic quantum electrodynamics,
we derive the complete result for the two-loop correction
to order $(\alpha/\pi)^2 (Z\,\alpha)^4$ for the $g$ factor of an electron
bound in an $nS$ state of a hydrogenlike ion. The results obtained
significantly improve the accuracy of the theoretical predictions
for the hydrogenlike carbon and oxygen ions and 
influence the value of the electron mass
inferred from $g$ factor measurements.
\end{abstract}


\maketitle

\section{Introduction}

The $g$ factor of a bound electron  is the coupling 
constant of the spin to an external, homogeneous magnetic field.
In natural units $\hbar = c = \varepsilon_0 = 1$,
it is defined by the relation
\begin{equation}
\delta E = -\frac{e}{2\,m}\,
\left< \vec{\sigma}\cdot \vec{B} \right> \,\frac{g}{2}\,,
\label{01}
\end{equation}
where $\delta E$ is the energy shift of the electron due to the
interaction with the magnetic field $\vec{B}$, 
$m$ is the mass of the electron, 
and $e$ is the physical electron charge ($e<0$).
The matrix $\vec{\sigma} \cdot \vec{B}$ contains the 
Pauli spin matrices $\vec{\sigma}$ and has eigenvalues $\pm |\vec B|$.

Studies of the free-electron $g$ factor play an important role
in modern physics. Together with the discovery of the Lamb shift in
hydrogen, the observation of the electron magnetic moment anomaly led
to the development of quantum electrodynamics (QED). 
After decades of intensive theoretical and
experimental studies, the free-electron $g$ factor provides
one of the most accurate and stringent tests of QED~\cite{hughes:99}. 
With the increased experimental and
theoretical precision, it presently yields the most accurate
determination of the fine-structure constant $\alpha$ \cite{mohr:05:rmp}.

It has not been until recently that investigations of the {\em bound}-electron
$g$ factor came into prominence. As was demonstrated in
Ref.~\cite{beier:02:prl}, the theoretical value of the
bound-electron $g$ factor can be used for the
determination of the mass of the electron when combined with an
experimental value for the ratio of
the electronic Larmor precession frequency $\omega_{\rm L}$ and
the cyclotron frequency of the ion in the trap $\omega_{\rm c}$,
\begin{equation}
m = m_{\rm ion}\,
\frac{g}{2}\, \frac{|e|}{q}\,
\frac{\omega_{\rm c}}{\omega_{\rm L}}
\,,
\label{02}
\end{equation}
where $q$ is the charge of the ion and $m_{\rm ion}$ is its mass.
The accuracy of the best experimental results for light hydrogenlike ions
\cite{haeffner:00:prl,verdu:04} is already below the 1 part
per billion level and is likely to be improved in the future. 
According to the recent adjustment of fundamental constants~\cite{mohr:05:rmp},
these measurements provide the most accurate method for the 
determination of the electron mass.

In order to match the experimental precision achieved,
various binding and QED corrections to the
bound-electron $g$ factor have to be calculated.
It has been found long ago~\cite{breit:28} that
in a relativistic (Dirac) theory,
the $g$ factor of a bound electron
differs from the value $g=2$ due to the so-called binding corrections.
For an $nS$ state, they are given by
\begin{align}
g^{(0)} &= \frac23 \left( 1+2\, \frac{E}{m}\right) 
\nonumber\\
&= 2-\frac{2}{3}\,\frac{(Z\,\alpha)^2}{n^2}+
\biggl(\frac{1}{2\,n}-\frac{2}{3}\biggr)\,\frac{(Z\,\alpha)^4}{n^3} +
\ldots \,,
\label{03}
\end{align}
where $E$ is the Dirac energy. Other corrections to
the bound-electron $g$ factor arise from the QED theory. They were
the subject of extensive theoretical investigations during the last
decade. Accurate calculations of the one-loop self-energy
\cite{blundell:97:pra,persson:97:g,beier:00:pra,yerokhin:02:prl,%
yerokhin:04:pra}, vacuum-polarization~\cite{persson:97:g,beier:00:pra,%
karshenboim:02:plb,lee:05:pra}, nuclear-recoil~\cite{shabaev:01:rec,%
martynenko:01:jetp,shabaev:02:prl}, and 
nuclear-polarizability~\cite{nefiodov:02:prl} 
corrections have been carried out. Detailed
$g$ factor investigations have been performed also for other systems
that could be of experimental interest in the near future,
in particular,
for Li-like ions \cite{glazov:04:pra} and hydrogenlike ions with a non-zero
nuclear spin \cite{moskovkin:04}.

The subject of this work is the two-loop QED correction,
which is presently the main source of the uncertainty of
theoretical predictions for the $g$ factor of hydrogenlike ions. We
present a complete calculation of this correction up to the 
order of $(\alpha/\pi)^2\,(Z\,\alpha)^4$. This two-loop correction has
already been addressed to in our former 
work~\cite{pachucki:04:prl}, where an incomplete calculation using a
photon-mass regularization was presented and an estimate for the
total contribution up to the order $(\alpha/\pi)^2\,(Z\,\alpha)^4$ was
obtained. The present computational method is based on the
dimensionally-regularized nonrelativistic quantum electrodynamics
(NRQED), which is a relatively new and very powerful approach for
the calculation of higher-order relativistic and radiative effects. It has
already been successfully applied to several challenging problems,
e.g., to the calculation of the positronium hyperfine
splitting~\cite{czarnecki:99:prl} and the ground-state energy of
the helium atom~\cite{korobov:01:prl}.

\section{Dimensionally regularized NRQED}
\label{sec:1}

As is customary in dimensionally regularized QED, we here 
assume that the dimension of the space-time 
is $D= 4-2\,\varepsilon$, and that of the space
$d=3-2\,\varepsilon$. The parameter $\varepsilon$ is considered as small,
but only on the level of matrix elements, where an analytic 
continuation to a noninteger spatial dimension is allowed.

Let us briefly discuss the extension of the basic formulas of NRQED to
the case of an arbitrary number of dimensions.
The momentum-space representation of the photon propagator preserves
its form, namely $g_{\mu\nu}/k^2$. The Coulomb interaction 
is~\cite{czarnecki:99:prl}
\begin{align}
V(r) &= -Z\,e^2\,\int\frac{d^d k}{(2\,\pi)^d}\,
\frac{e^{{\rm i}\,\vec k\cdot \vec r}}{k^2} 
\nonumber\\
&= -\frac{Z\,e^2}{4\,\pi\,r^{1-2\,\varepsilon}}\,
\left[(4\,\pi)^\varepsilon\,
\frac{\Gamma(1-2\,\varepsilon)}{\Gamma(1-\varepsilon)}\right]
\equiv -\frac{Z_\varepsilon\,\alpha}{r^{1-2\,\varepsilon}}\,,
\label{04}
\end{align}
where the latter representation provides an implicit definition of 
$Z_\varepsilon$, and we have
used the formula for the surface area of a $d$-dimensional unit sphere
\begin{equation}
\Omega_d = \frac{2\,\pi^{d/2}}{\Gamma(d/2)}\,.
\label{05}
\end{equation}
The nonrelativistic Hamiltonian of the hydrogenic system is
\begin{equation}
H_0 = \frac{\vec{p}^{\;2}}{2\,m}
-\frac{Z_\varepsilon\,\alpha}{r^{1-2\,\varepsilon}}\,.
\label{06}
\end{equation}
The operator $\vec{p}^{\;2}$ is well defined in any integer
dimension. If we restrict our consideration to the spherically symmetric states,
$\vec{p}^{\;2}$ can be continued to an arbitrary real dimension by
\begin{equation}
\vec{p}^{\;2} = -\frac{1}{r^{d-1}}\,
\frac{\partial}{\partial r}\,r^{d-1}\,\frac{\partial}{\partial r}\,.
\label{07}
\end{equation}
In the following, we will not need the explicit (unknown) form of the solution 
of the Schr\"odinger equation in $d$ dimensions. It will
be sufficient to use instead its scaling properties,
which we obtain by introducing the dimensionless radial
variable $\rho$
\begin{equation}
\rho = (m\,\alpha)^{\frac{1}{1+2\,\varepsilon}}\,r\,.
\label{08}
\end{equation}
In atomic units, i.e.~expressed as a 
function of the dimensionless $\rho$, 
the Schr\"odinger Hamiltonian takes the form
\begin{equation}
H_0 =  \alpha^{\frac{2}{1+2\,\varepsilon}}\,
m^{\frac{1-2\,\varepsilon}{1+2\,\varepsilon}}\,
\biggl(\frac{\vec p_\rho^{\;2}}{2}
-\frac{Z_\varepsilon}{\rho^{1-2\,\varepsilon}}\biggr)\,.
\label{09}
\end{equation}

We now turn to relativistic corrections to the Schr\"odinger
Hamiltonian in an arbitrary number of dimensions.
These corrections can be obtained from the Dirac Hamiltonian
by the Foldy-Wouthuysen transformation.
In order to incorporate a part of radiative effects right from the
beginning, we use an effective Dirac Hamiltonian modified by the
electromagnetic form factors $F_1$ and $F_2$
(see, e.g., Ch.~7 of~\cite{itzykson:80})
 \begin{align}
H =& \vec{\alpha} \cdot
\left[\vec{p} - e \, F_1(\vec \nabla^2) \, \vec{A}\right] + \beta\,m 
+ e\,F_1(\vec \nabla^2) \, A_0
\nonumber\\
& + F_2(\vec \nabla^2) \, \frac{e}{2\,m} \, \left({\rm i}\,\vec{\gamma} \cdot
\vec{E} - \frac{\beta}{2} \, \Sigma^{ij}\,B^{ij} \right)\,,
\label{10}
\end{align}
where
\begin{eqnarray}
B^{ij} &=& \nabla^i\,A^j-\nabla^j\,A^i\,,\label{11}\\
\Sigma^{ij} &=& \frac{\rm i}{2}\,[\gamma^i,\gamma^j]\,.
\label{12}
\end{eqnarray}
We use three-dimensional notations here, namely 
$\nabla^i \equiv \partial_i = \partial/\partial x^i$.
Formulas for the electromagnetic
form factors $F_{1,2}$ can be found in Appendix A.

Having the Foldy-Wouthuysen transformation defined by the operator $S$
(see Ref.~\cite{pachucki:04:lwqed}, and $\kappa \equiv F_2(0)$)
\begin{align}
S =& -\frac{\rm i}{2\,m}\,\left\{
\beta\,\vec\alpha\cdot\vec\pi
-\frac{1}{3\,m^2}\,\beta\,(\vec\alpha\cdot\vec\pi)^3\right.
\nonumber\\
& \left. +\frac{e(1+\kappa)}{2\,m}\,{\rm i}\,\vec\alpha\cdot\vec E
-\frac{e\,\kappa}{8\,m^2}\,
[\vec\alpha\cdot\vec\pi, \beta\,\Sigma^{ij}\,B^{ij}]\right\}\,,
\label{13}
\end{align}
the new Hamiltonian is obtained via 
\begin{subequations}
\begin{equation}
\label{trafo}
H' = e^{{\rm i}\,S}\,(H-i\,\partial_t)\,e^{-{\rm i}\,S}
\end{equation}
and takes the form
\begin{align}
H' =& \frac{\vec \pi^{\;2}}{2\,m} + 
e\,[1 + {F'}_1(0)\,{\vec \nabla}^2] A^0 -
\frac{e}{4\,m}\,(1+\kappa)\,\sigma^{ij}\,B^{ij}
\nonumber\\
& -\frac{\vec \pi^{\;4}}{8\,m^3}
-\frac{e}{8\,m^2}\,(1+2\,\kappa)\,
\left[\vec\nabla\cdot\vec E+\sigma^{ij}\,\{E^i,\pi^j\}\right]
\nonumber \\ 
& +\frac{e}{8\,m^3}\left[(1+\kappa)\,p^2\,\sigma^{ij}\,B^{ij}+
2\,\kappa\,p^k\,\sigma^{ki}\,B^{ij}\,p^j\right]
\nonumber \\ 
& -\frac{e}{8\,m^2}\,[{F'}_1(0)+2\,{F'}_2(0)]\,\sigma^{ij}\,
\{\vec \nabla^2 E^i,\pi^j\}
+\ldots
\label{14}\,,
\end{align}
\end{subequations}
where by dots we denote the omitted higher-order terms,
$\{X,Y\} \equiv X\,Y+Y\,X$, and 
$\sigma^{ij} = [\sigma^i,\,\sigma^j]/(2\,{\rm i})$.
The Hamiltonian $H'$ is a generalization of the 
Foldy-Wouthuysen Hamiltonian $H_{FW}$~\cite{pachucki:04:lwqed}
to an arbitrary number of dimensions.
The electromagnetic field in $H'$
is the sum of the external Coulomb field, 
the external (constant) magnetic field,
and a slowly varying field of the radiation.
For practical calculations, it is more convenient to
have a Hamiltonian expressed in terms of the gauge-independent field
strengths. To achieve this, we separate out the Coulomb field
and perform the Power-Zienau transformation of the Hamiltonian $H'$  
with the operator $S'$ of the form \cite{pachucki:04:lwqed}
\begin{equation}
S' = -e\int_0^1 du\,\vec r\cdot\vec A(u\,\vec r,t)\,.
\label{15}
\end{equation}
After neglecting irrelevant spin-independent terms, 
the transformed Hamiltonian becomes
\begin{widetext}
\begin{align}
H'' =& \frac{p^2}{2\,m}+V -e\,\vec r\cdot\vec E
+\frac{(1+2\,\kappa)}{8\,m^2}\,
\frac{V'}{r}\,\sigma^{ij}\,L^{ij}
-\frac{e}{4\,m}\,\left[L^{ij}+(1+\kappa)\,\sigma^{ij}\right]\,B^{ij}
\nonumber \\ 
& +\frac{e}{8\,m^3}\left[(1+\kappa)\,p^2\,\sigma^{ij}\,B^{ij}+
2\,\kappa\,p^k\,\sigma^{ki}\,B^{ij}\,p^j\right]
-\frac{e\,(1+2\,\kappa)}{8\,m^2}\,\frac{V'}{r}\,
\sigma^{ij}\,r^j\,B^{ik}\,r^k
\nonumber \\ 
& +\frac{e^2\,(1+2\,\kappa)}{8\,m^2}\,\sigma^{ij}\,E^j\,B^{ik}\,r^k
-\frac{e\,(1+\kappa)}{4\,m}\,\sigma^{ij}\,r^k\,B^{ij}_{,k}
-\frac{e\,(1+2\,k)}{4\,m^2}\,\sigma^{ij}\,E^i\,p^j
\nonumber \\ 
& +{F'}_1(0)\,4\,\pi\,Z\,\alpha\,\delta^d(r)
-\frac{e}{8\,m^2}\,[{F'}_1(0)+2\,{F'}_2(0)]\,\sigma^{ij}\,
\nabla^j[4\,\pi\,Z\,\alpha\,\delta^d(r)]\,B^{ik}\,r^k\,.
\label{16}
\end{align}
\end{widetext}
Here, $L^{ij} = r^i\,p^j-r^j\,p^i$ and $B_{,k} \equiv \nabla^k B$.
$H''$ is the generalization of the Power-Zienau Hamiltonian 
$H_{PZ}$~\cite{pachucki:04:lwqed} to an arbitrary number of dimensions.

The Hamiltonian $H''$ includes most of the radiative corrections
that are needed for our calculation, but not all of them. 
First, the higher-order terms with the anomalous
magnetic moment are omitted in $H''$.
This contribution is more conveniently calculated
with the exact Dirac-Coulomb wave functions, starting directly
from the Hamiltonian (\ref{10}).
Furthermore, there is an additional correction that cannot be accounted for by
the $F_1$ and $F_2$ form factors. It is represented
by an effective local operator
that is quadratic in the field strengths. This operator is derived separately
by evaluating a low-energy limit of the electron scattering amplitude off
the Coulomb and the magnetic fields. Details of
this calculation are presented in Appendix B. The result is
\begin{equation}
\delta H =
\frac{e^2}{2\,m}\,\left[2\,\sigma^{ij}\,B^{ik}\,\nabla^j E^k\,\eta +
\sigma^{ij} B^{ij}\,\nabla^k E^k\,\xi\right]\,,
\label{17}
\end{equation}
where $B=$ const, $E$ is an arbitrary electric field, and the functions
$\eta$ and $\xi$ are given by Eqs.~(\ref{eta}) and (\ref{xi}),
respectively. 


\section{One-loop self-energy correction}

The dimensionally regularized NRQED approach formulated in
the previous section will be first employed for a derivation of the
self-energy correction to order $(\alpha/\pi)\,(Z\,\alpha)^4$ for the
bound-electron $g$ factor. This derivation will serve us as a test of the
new approach (as this result has been already obtained in our previous
work \cite{pachucki:04:prl}) and also as a basis for the
two-loop calculation.

As in~\cite{pachucki:04:prl}, we separate 
the one-loop self-energy correction up to the order 
of $(\alpha/\pi)\,(Z\,\alpha)^4$ into three parts,
\begin{equation}
g^{(1)} = g^{(1)}_1 +g^{(1)}_2 +g^{(1)}_3\,, \label{19}
\end{equation}
where the first part is the the contribution due to the
free-electron form factors $F_1$ and $F_2$, the second part is the
contribution induced by the additional Hamiltonian (\ref{17}), and
the third part is the contribution coming from low-energy 
photons, i.e. a  Bethe-logarithm type contribution. 

We start with the form-factor part $g^{(1)}_1$.
The anomalous magnetic moment $F_2(0)$ in the modified 
Dirac-Coulomb Hamiltonian (\ref{10}) leads to the following energy 
shift linear in the magnetic field ($d=3$)
\begin{align}
& \delta E_{1A} = \left< -F^{(1)}_2(0) \, \frac{e}{2\,m} \,
\beta \,\vec\Sigma\cdot\,\vec B\right>
\nonumber\\
& +2\,\left< F^{(1)}_2(0) \, \frac{{\rm i}\,e}{2\,m} \, 
\vec{\gamma} \cdot\vec{E}\,
\frac{1}{(E-H)'}\,(-e)\,\vec\alpha\cdot\vec A\right>\,,
\label{20}
\end{align}
where $\vec A = (\vec B\times\vec r\,)/2$,
and we denote the one-loop components of the 
form factors by the corresponding superscript.
The corresponding correction to the $g$ factor is
\begin{equation}
g^{(1)}_{1A} = 2\,F_2^{(1)}(0)\,\biggl[1+\frac{(Z\,\alpha)^2}{6\,n^2}+
\biggl(\frac{3}{2}-\frac{5}{24\,n}\biggr)\,\frac{(Z\,\alpha)^4}{n^3}\biggr]\,.
\label{21}
\end{equation}
In obtaining this result, we used the closed-form expression 
\cite{shabaev:91:jpb,shabaev:03:PSAS} for the
component of the Dirac wave function perturbed by the
magnetic interaction, which has the same relativistic angular momentum,
as the reference state.

For the remaining part of the form-factor contribution,
we employ the transformed Hamiltonian (\ref{16}).
The last term of this Hamiltonian
\begin{equation}
-\frac{e}{8\,m^2}\,[{F'}_1(0)+2\,{F'}_2(0)]\,\sigma^{ij}\,
\nabla^j[4\,\pi\,Z\,\alpha\,\delta^d(r)]\,B^{ik}\,r^k
\label{22}
\end{equation}
gives rise to a contribution
\begin{equation}
g^{(1)}_{1B} = -[{F'}_1^{(1)}(0) + 2\,{F'}_2^{(1)}(0)]\,
\langle 4\,\pi\,Z\,\alpha\,\delta^d(r)\rangle\,. \label{23}
\end{equation}
The second-order correction to the energy
\begin{equation}
2\,\left< {F'}_1^{(1)}(0)\,4\,\pi\,Z\,\alpha\,\delta^d(r)\,
\frac{1}{(E_0-H_0)'}\,
\frac{e}{8\,m^3}\,p^2\,\sigma^{ij}\,B^{ij}\right> \label{24}
\end{equation}
yields
\begin{equation}
g^{(1)}_{1C} = 2\,(3-8\,\varepsilon)\, {F'}_1^{(1)}(0)\,
\langle 4\,\pi\,Z\,\alpha\,\delta^d(r)\rangle\,.
\label{25}
\end{equation}
The other second-order correction to the energy
\begin{equation}
2\,\left< {F'}_1^{(1)}(0)\,\vec\nabla^2 V\,
\frac{1}{(E_0-H_0)'}\,
(-1)\frac{e}{8\,m^2}\,\frac{V'}{r}\,
\sigma^{ij}\,r^j\,B^{ik}\,r^k\right>,
\label{26}
\end{equation}
gives
\begin{equation}
g^{(1)}_{1D} = -(1-4\,\varepsilon)\, {F'}_1^{(1)}(0)\,\langle 4\,\pi\,Z\,\alpha\,\delta^d(r)\rangle\,.
\label{27}
\end{equation}
The total form-factor contribution is
\begin{eqnarray}
g^{(1)}_{1}  &=& g^{(1)}_{1A} + g^{(1)}_{1B} + g^{(1)}_{1C} + g^{(1)}_{1D}\,. \label{28}
\end{eqnarray}

The second part of Eq.~(\ref{19}), denoted by $g^{(1)}_2$, 
is a high-energy correction that is not accounted for by
the form factors. It is given by the effective Hamiltonian (\ref{17}),
with $E$ being the electric Coulomb field.
The corresponding correction to the $g$ factor is
\begin{align}
g^{(1)}_{2} &= 4\biggl(\frac{2}{d}\,\eta^{(1)} + \xi^{(1)}\biggr)\,
\langle 4\,\pi\,Z\,\alpha\,\delta^d(r)\rangle 
\nonumber\\
&= \frac{\alpha}{\pi}\,\biggl(\frac{2}{9\,\varepsilon} 
+ \frac{19}{27} \biggr)\,\langle 4\,\pi\,Z\,\alpha\,\delta^d(r)\rangle\,,
\label{29}
\end{align}
where $\eta^{(1)}$ and $\xi^{(1)}$ are the one-loop components
of the coefficient functions given below in Eqs.~(\ref{eta})
and~(\ref{xi}).

The third part of Eq.~(\ref{19}) is a low-energy contribution 
that can be considered
as a correction to the Bethe logarithm due to the interaction with an
external magnetic field. Let us first derive
the Bethe-logarithm correction to the hydrogen Lamb shift within
the dimensional regularization. The correction to the energy is
\begin{align}
\delta E_L =& e^2\int\frac{d^dk}{(2\,\pi)^d\,2\,k}\,
\delta^{ij}_T\,\left<
\frac{p^i}{m}\,\frac{1}{E_0-k-H_0}\,\frac{p^j}{m}\right>
\nonumber \\
=& e^2\int\frac{d^dk}{(2\,\pi)^d\,2\,k}\,
\delta^{ij}_T\,k^2\,\left<
r^i\,\frac{1}{E_0-k-H_0}\,r^j\right>\,.
\label{30}
\end{align}
Here, $\delta^{ij}_T = \delta^{ij}-k^i\,k^j/k^2$
is the transverse delta function, and 
$k = |\vec k|$. After performing the integration over $k$ and dropping a 
common overall factor of
$(4\,\pi)^\varepsilon\,\Gamma(1+\varepsilon)$,
$\delta E_L$ becomes
\begin{align}
\delta E_L =& \frac{\alpha}{\pi}\,\frac{1}{6\,\varepsilon}\,
\frac{\langle 4\,\pi\,Z\,\alpha\,\delta^d(r)\rangle}{m^2}
\nonumber\\
& +m\,\frac{\alpha}{\pi}\,\frac{(Z\,\alpha)^4}{n^3}\,
\left[\frac{10}{9}-\frac{4}{3}\,\ln(Z\,\alpha)^2]-
\frac{4}{3}\,\ln k_0\right]\,,
\label{31}
\end{align}
where the Bethe logarithm $\ln k_0$ is given by
\begin{equation}
\ln k_0 = \frac{\left<\vec
  p\,(H_0-E_0)\ln\left[\frac{2\,(H_0-E_0)}{m\,(Z\,\alpha)^2}\right]
\,\vec p\right>}{\langle \vec p\,(H_0-E_0)\,\vec p\rangle}\,.
\label{32}
\end{equation}
We now consider all corrections to $\delta E_L$
due to the presence of the external magnetic field.
The first one is induced by the correction to the Hamiltonian
[the fifth term on the right-hand side of Eq.~(\ref{16})]
\begin{equation}
\delta_A H = \frac{p^2}{8\,m^3}\,e\,\sigma^{ij}\,B^{ij}\,.\label{33}
\end{equation}
The corresponding energy shift is given by
\begin{equation}
\delta_A E =  e^2\int\frac{d^dk}{(2\,\pi)^d\,2\,k}\,
\delta^{ij}_T\,k^2\,
\delta_A\left<r^i\,\frac{1}{E_0-k-H_0}\,r^j\right>\,,\label{34}
\end{equation}
where by $\delta_A\langle\ldots\rangle$ we denote the 
first-order correction to the matrix
element induced by the perturbing Hamiltonian $\delta_A H$. 
This matrix element
is calculated using the scaling properties of the Schr\"odinger
Hamiltonian given by Eq. (\ref{09}), and
the corresponding correction to the $g$ factor is found to be
\begin{align}
g^{(1)}_{3A} =& \frac{\alpha}{\pi}\,\frac{1}{3\,\varepsilon}\,
\langle 4\,\pi\,Z\,\alpha\,\delta^d(r)\rangle
\nonumber\\
& -\frac{\alpha}{\pi}\,\frac{(Z\,\alpha)^4}{n^3}\,
\left[\frac{8}{3}\,\ln(Z\,\alpha)^2
+\frac{8}{3}\,\ln k_0 +\frac{100}{9}\right]\,.
\label{35}
\end{align}
The second correction to the interaction with the magnetic field is
[sixth term in Eq.~(\ref{16})]
\begin{align}
\delta_B H =& 
-\frac{e}{8\,m^2}\,\frac{V'}{r}\, \sigma^{ij}\,r^j\,B^{ik}\,r^k 
\nonumber\\
=& -\frac{d-2}{d}\,\frac{e}{8\,m^2}\,V\,\sigma^{ij}\,B^{ij}\,,
\label{36}
\end{align}
where the last part of the equation holds only for S-states.
The corresponding contribution to the $g$ factor is
\begin{align}
& g^{(1)}_{3B} = -\frac{\alpha}{\pi}\,\frac{2}{9\,\varepsilon}\,
\langle 4\,\pi\,Z\,\alpha\,\delta^d(r)\rangle
\nonumber\\
& +\frac{\alpha}{\pi}\,\frac{(Z\,\alpha)^4}{n^3}\,
\left[\frac{16}{9}\,\ln(Z\,\alpha)^2
+\frac{16}{9}\,\ln k_0 +\frac{64}{27}\right]\,.
\label{37}
\end{align}
The third correction is due to the coupling with the radiation field,
[seventh term in Eq.~(\ref{16})],
\begin{equation}
\delta_C H = \frac{e^2}{8\,m^2}\,\sigma^{ij}\,E^j\,B^{ik}\,r^k 
= [-e\,\vec r\cdot\vec E]\,
\left[\frac{-e\,\sigma^{ij}\,B^{ij}}{8\,m^2\,d}\right]\,.
\label{38}
\end{equation}
Here, the last expression is obtained by $d$-dimensional angular averaging.
The corresponding energy shift is written as
\begin{align}
\delta_C E &= 2 \left[\frac{-e\,\sigma^{ij}\,B^{ij}}{8\,m^2\,d}\right]
\nonumber \\ &\times
e^2\int\frac{d^dk}{(2\,\pi)^d\,2\,k}\,\delta^{ij}_T\,k^2\,\bigl\langle
r^i\,\frac{1}{E_0-k-H_0}\,r^j\bigr\rangle\,.
\label{39}
\end{align}
The contribution to the $g$ factor is
\begin{align}
g^{(1)}_{3C} =& \frac{\alpha}{\pi}\,\frac{1}{9\,\varepsilon}\,
\langle 4\,\pi\,Z\,\alpha\,\delta^d(r)\rangle
\nonumber\\
& -\frac{\alpha}{\pi}\,\frac{(Z\,\alpha)^4}{n^3}\,
\left[\frac{8}{9}\,\ln(Z\,\alpha)^2
+\frac{8}{9}\,\ln k_0 -\frac{28}{27}\right]\,.
\label{40}
\end{align}
The fourth contribution involves both the correction to the coupling
with the radiation field and the interaction with the magnetic field,
[the fourth and the ninth term of Eq.~(\ref{16})],
\begin{equation}
\delta_D H = 
-\frac{e}{4\,m^2}\,\sigma^{ij}\,E^i\,p^j -\frac{e}{4\,m}\,L^{ij}\,B^{ij}
\label{41}
\end{equation}
and is of the form
\begin{align}
& \delta_D E = 2\,e^2\int\frac{d^dk}{(2\,\pi)^d\,2\,k}\,\delta^{ij}_T\,k^2\,
\nonumber \\ 
& \times \left<
r^i\frac{1}{E_0-k-H_0}\left[\frac{-e}{4\,m} L^{ab} B^{ab}\right]
\frac{1}{E_0-k-H_0}\frac{\sigma^{jk}\,p^k}{4\,m^2}\right>.
\label{42}
\end{align}
The corresponding correction to the $g$ factor is
\begin{align}
g^{(1)}_{3D} =& \frac{\alpha}{\pi}\,\frac{1}{3\,\varepsilon}\,
\langle 4\,\pi\,Z\,\alpha\,\delta^d(r)\rangle
\nonumber\\
& -\frac{\alpha}{\pi}\,\frac{(Z\,\alpha)^4}{n^3}\,
\left[\frac{8}{3}\,\ln(Z\,\alpha)^2
+\frac{8}{3}\,\ln k_0 -\frac{20}{9}\right]\,.
\label{43}
\end{align}
The fifth contribution is due to another correction to the coupling
with the radiation field and
the same interaction with the magnetic field,
\begin{equation}
\delta_E H =  -\frac{e}{4\,m}\,\sigma^{ij}\,r^k\,B^{ij}_{,k}
-\frac{e}{4\,m}\,L^{ij}\,B^{ij}\,,
\label{44}
\end{equation}
and is of the form
\begin{widetext}
\begin{align}
\delta_E E = \frac{4\,e^2}{d}\,\int\frac{d^dk}{(2\,\pi)^d\,2\,k}\,k^2\,
\left< r^i \frac{1}{E_0-k-H_0} \left[-\frac{e}{4\,m} L^{ab}\,B^{ab}\right]
\frac{1}{E_0-k-H_0} \frac{{\rm i}\,k\,\sigma^{ij}r^j}{4\,m^2}\right>\,.
\label{45}
\end{align}
The corresponding correction to the $g$ factor is
\begin{align}
g^{(1)}_{3E} = -\frac{\alpha}{\pi}\,\frac{4}{9\,\varepsilon}\,
\langle 4\,\pi\,Z\,\alpha\,\delta^d(r)\rangle
+\frac{\alpha}{\pi}\,\frac{(Z\,\alpha)^4}{n^3}\,
\left[\frac{32}{9}\,\ln(Z\,\alpha)^2+\frac{32}{9}\,\ln k_0
  -\frac{136}{27}\right]\,.
\label{46}
\end{align}
The sixth and the last contribution is due to the spin-orbit interaction and
the interaction to the magnetic field,
\begin{equation}
\delta_F H = \frac{1}{8\,m^2}\,\frac{V'}{r}\,\sigma^{ij}\,L^{ij}
-\frac{e}{4\,m}\,L^{ij}\,B^{ij}\,.
\label{47}
\end{equation}
This correction involves a more complicated  matrix element with three
propagators,
\begin{align}
\delta_F E = 2\,e^2\int\frac{d^dk}{(2\,\pi)^d\,2\,k}\,
\delta^{ij}_T\,k^2\,
\left< r^i\,\frac{1}{E_0-k-H_0}\, 
%
\left[\frac{1}{8\,m^2}\,\frac{V'}{r}\,
\sigma^{ij}\,L^{ij}\right] \, \frac{1}{E_0-k-H_0}
\left[-\frac{e}{4\,m}\,L^{ij}\,B^{ij}\right]\,\frac{1}{E_0-k-H_0}\,
r^j\right>\,.
\label{48}
\end{align}
\end{widetext}
The corresponding correction to the $g$ factor is
\begin{align}
g^{(1)}_{3F} =& \frac{\alpha}{\pi}\,\frac{1}{3\,\varepsilon}\,
\langle 4\,\pi\,Z\,\alpha\,\delta^d(r)\rangle
\nonumber\\
& -\frac{\alpha}{\pi}\,\frac{(Z\,\alpha)^4}{n^3}\,
\left[\frac{8}{3}\,\ln(Z\,\alpha)^2
+\frac{8}{3}\,\ln k_3 -\frac{20}{9}\right]\,,
\label{49}
\end{align}
where $\ln k_3$ is implicitly defined by the relation
\begin{align}
& \int_0^\epsilon d k\, k^2\,
\left\langle\vec{r}\,\frac{1}{E_0-H_0-k}\,
\frac{1}{r^3}\,\frac{1}{E_0-H_0-k}\,\vec{r}\right\rangle
\nonumber\\
& \quad = \epsilon\,\left<\frac{1}{r}\right>
-4\,\frac{(Z\,\alpha)^3}{n^3} \left[\ln \frac{2\,\epsilon}{(Z\,\alpha)^2}
-\ln k_3\right],\label{50}
\end{align}
which holds in the limit of large $\epsilon$.

Finally, the total Bethe-logarithm type contribution 
to the $g$ factor is a sum of calculated terms
\begin{equation}
g^{(1)}_{3} = g^{(1)}_{3A} + g^{(1)}_{3B} + g^{(1)}_{3C} +
              g^{(1)}_{3D} + g^{(1)}_{3E} + g^{(1)}_{3F}\,. \label{51}
\end{equation}
The complete one-loop self-energy correction to the bound-electron $g$
factor is then
\begin{align}
& g^{(1)} = \frac{\alpha}{\pi} \left\{1 + \frac{(Z\alpha)^2 }{6 n^2}
+ \frac{(Z\alpha)^4}{n^3} \left[\frac{32}{9} \ln[(Z\alpha)^{-2}] 
\right.\right.
\nonumber\\
& \quad \left. \left.
+ \frac{73}{54} - \frac{5}{24 n}
- \frac{8}{9} \ln k_0 - \frac{8}{3} \ln k_3 \right] \right\}\,,
\label{52}
\end{align}
in full agreement with the former result in 
Eq.~(12) of Ref.~\cite{pachucki:04:prl}.

\section{Two-loop contribution}

The derivation of the two-loop corrections to the bound-electron $g$ factor is
performed in full analogy to the one-loop calculations.
The total two-loop correction of the order $(\alpha/\pi)^2\,(Z\,\alpha)^4$
can be separated into four parts,
\begin{equation}
g^{(2)} = g^{(2)}_1 +g^{(2)}_2 +g^{(2)}_3 +g^{(2)}_4\,.
\label{53}
\end{equation}
The first part $g^{(2)}_1$ is a form-factor contribution.
The second part $g^{(2)}_2$ is an 
additional high-energy contribution not accounted for by the form factors.
The third part arises from a contribution
in which one of the two virtual photons is of low energy. The second
photon effectively modifies the vertex, which can be accounted for by
the anomalous magnetic moment.
The fourth contribution $g^{(2)}_4$ involves the closed fermion loops
and is called the vacuum polarization part.

We start with the form-factor contribution. The two-loop anomalous magnetic
moment correction is obtained from the corresponding one-loop
contribution, Eq.~(\ref{21})
\begin{equation}
g^{(2)}_{1A} = 2\,F_2^{(2)}(0)\,\left[1+\frac{(Z\,\alpha)^2}{6\,n^2}+
\biggl(\frac{3}{2}-\frac{5}{24\,n}\biggr)\,\frac{(Z\,\alpha)^4}{n^3}\right]\,.
\label{54}
\end{equation}
The one-loop anomalous magnetic moment in the modified Dirac-Coulomb
Hamiltonian (\ref{10}) leads to the following energy shift,
%
\begin{align}
& \delta E_{1B} = [F^{(1)}_2(0)]^2\,\left\{\left< \frac{-e}{2\,m} \,
\beta \,\vec\Sigma\cdot\,\vec B \,\frac{1}{(E-H)'}\,
\frac{{\rm i}\,e}{2\,m} \, \vec{\gamma}\right.
\cdot\vec{E}\right>
\nonumber \\ &
+2\left< \frac{{\rm i}\,e}{2\,m} \vec{\gamma}\cdot\vec{E}
\frac{1}{(E-H)'} \frac{{\rm i}\,e}{2\,m} \vec{\gamma}\cdot\vec{E}
\frac{1}{(E-H)'} (-e)\,\vec\alpha\cdot\vec A\right>
\nonumber \\ &
+\left< \frac{{\rm i}\,e}{2\,m} \,\vec{\gamma}\cdot\vec{E}\,
\frac{1}{(E-H)'}\,(-e)\,\vec\alpha\cdot\vec A\,
\frac{1}{(E-H)'}\,\frac{{\rm i}\,e}{2\,m} \,\vec{\gamma}\cdot\vec{E}\,\right>
\nonumber \\ &
-\left< -e\,\vec\alpha\cdot\vec A\right>\,
\left< \frac{{\rm i}\,e}{2\,m} \,\vec{\gamma}\cdot\vec{E}\,
\frac{1}{(E-H)'^{\,2}}\,\frac{{\rm i}\,e}{2\,m} \,
\vec{\gamma}\cdot\vec{E}\right>
\nonumber \\ &
\left.
-2\,\left<\frac{{\rm i}\,e}{2\,m} \,\vec{\gamma}\cdot\vec{E}\right>\,
\left< (-e)\,\vec\alpha\cdot\vec A\,
\frac{1}{(E-H)'^{\,2}}\,\frac{{\rm i}\,e}{2\,m} \,
\vec{\gamma}\cdot\vec{E}\right>
\right\}\,.
\label{55}
\end{align}
%
The corresponding correction to the $g$ factor is
\begin{equation}
g^{(2)}_{1B} =
-\frac{2}{3}\,[F_2^{(1)}(0)]^2\,\frac{(Z\,\alpha)^4}{n^3}\,.
\label{56}
\end{equation}
The other contributions due to the two-loop form factors are immediately
obtained from the corresponding one-loop expressions 
in Eqs.~(\ref{23}),~(\ref{25}) and~(\ref{27})
\begin{align}
g^{(2)}_{1C} =& -[{F'}_1^{(2)}(0) + 2\,{F'}_2^{(2)}(0)]\,
\langle 4\,\pi\,Z\,\alpha\,\delta^d(r)\rangle \,, \label{57}\\
g^{(2)}_{1D} =& 2\,(3-8\,\varepsilon)\, {F'}_1^{(2)}(0)\,
\langle 4\,\pi\,Z\,\alpha\,\delta^d(r)\rangle \,,\label{58}\\
g^{(2)}_{1E} =& -(1-4\,\varepsilon)\, {F'}_1^{(2)}(0)\,\langle
4\,\pi\,Z\,\alpha\,\delta^d(r)\rangle
\,.
\label{59}
\end{align}
The second-order corrections involving the slope of the one-loop
form factors and the one-loop anomalous magnetic moment vanish. It
becomes clear if we notice that the coupling of the anomalous
magnetic moment to the magnetic field, as obtained from the
Hamiltonian (\ref{16}), is
\begin{align}
\delta V =& \frac{e\,\kappa}{8\,m^3}
\left[p^2\,\sigma^{ij}\,B^{ij}+
2\,p^k\,\sigma^{ki}\,B^{ij}\,p^j
\right.
\nonumber\\
& \left. -2\,m\,\frac{V'}{r}\,
\sigma^{ij}\,r^j\,B^{ik}\,r^k \right]\label{60}
\end{align}
and for $S$-states
\begin{equation}
\delta V = \frac{d-2}{d}\,\frac{e\,\kappa}{4\,m^2}\,\sigma^{ij}\,B^{ij}\,
\left[\frac{\vec p^{\;2}}{2\,m}-\frac{Z_\epsilon\,\alpha}
{r^{1-2\,\epsilon}}\right]\,.\label{61}
\end{equation}
All other possible two-loop corrections, which involve
one-loop form-factors, are of higher order in the $Z\,\alpha$ expansion. 
Therefore, the total form-factor contribution is given by the sum
\begin{eqnarray}
g^{(2)}_{1}  &=& g^{(2)}_{1A} + g^{(2)}_{1B} + g^{(2)}_{1C} +
g^{(2)}_{1D}+ g^{(2)}_{1E}
\,.\label{62}
\end{eqnarray}

The second part of Eq.~(\ref{53}) is a high-energy correction 
that is not accounted for by
the form factors.  This contribution is induced by the effective
Hamiltonian $\delta H$ in Eq. (\ref{17}),
with $E$ being the electric Coulomb field.
The corresponding correction to the $g$ factor is
\begin{align}
g^{(2)}_{2} 
=& 4\biggl(\frac{2}{d}\,\eta^{(2)} + \xi^{(2)}\biggr)\,
\langle 4\,\pi\,Z\,\alpha\,\delta^d(r)\rangle
\nonumber \\ 
=&
\biggl(- \frac{5}{9\,\varepsilon}
+ \frac{5455}{972}  
+ \frac{833}{1296}\,{\pi }^2 
- \frac{31}{9}\,{\pi }^2\,\ln 2 
+ \frac{31}{6}\,\zeta(3)\biggr)
\nonumber \\ 
& \times \langle 4\,\pi\,Z\,\alpha\,\delta^d(r)\rangle \,.\label{63}
\end{align}

The third part of Eq.~(\ref{53}) $g^{(2)}_{3}$
is obtained from the formulas for the one-loop Bethe-logarithm
corrections. The overall coefficients in these formulas 
are modified by the presence of the anomalous magnetic moment $\kappa$,
in accordance with the corresponding terms in the effective Hamiltonian (16). 
The resulting corrections to the Hamiltonian describing 
the interaction with the magnetic field are
given by (for S-states)
\begin{subequations}
\label{69}
\begin{align}
\delta^{(2)}_A H =& \frac{e\,\kappa}{8\,m^3}\bigl[p^2\,\sigma^{ij}\,B^{ij}+
              2\,p^k\,\sigma^{ki}\,B^{ij}\,p^j\bigr] 
\nonumber\\
=& \left[\kappa\,\frac{(d-2)}{d}\right]\,
   \frac{p^2}{8\,m^3}\,e\,\sigma^{ij}\,B^{ij}\,,\\
\delta^{(2)}_B H = & [2\,\kappa]\,\biggl(
  -\frac{d-2}{d}\,\frac{e}{8\,m^2}\,V\,\sigma^{ij}\,B^{ij}\biggr)\,,\\
\delta^{(2)}_C H = & 
  [2\,\kappa]\,\frac{e^2}{8\,m^2}\,\sigma^{ij}\,E^j\,B^{ik}\,r^k\,,\\
\delta^{(2)}_D H = & 
  [2\,\kappa]\,\biggl(-\frac{e}{4\,m^2}\,\sigma^{ij}\,E^i\,p^j\biggr)
                -\frac{e}{4\,m}\,L^{ij}\,B^{ij} \,,\\
\delta^{(2)}_E H = & 
  [\kappa]\,\biggl(-\frac{e}{4\,m}\,\sigma^{ij}\,r^k\,B^{ij}_{,k}\biggr)
                -\frac{e}{4\,m}\,L^{ij}\,B^{ij} \,,\\
\delta^{(2)}_F H = & 
  [2\,\kappa]\,\frac{1}{8\,m^2}\,\frac{V'}{r}\,\sigma^{ij}\,L^{ij}
                -\frac{e}{4\,m}\,L^{ij}\,B^{ij}\,.
\end{align}
\end{subequations}
The resulting two-loop corrections to the $g$ factor are
\begin{subequations}
\label{70}
\begin{eqnarray}
g^{(2)}_{3A} &=& \left[\kappa\,\frac{(d-2)}{d}\right]\, g^{(1)}_{3A}\,, \\
g^{(2)}_{3B} &=& 2\,\kappa\,g^{(1)}_{3B}\,, \\
g^{(2)}_{3C} &=& 2\,\kappa\,g^{(1)}_{3C}\,, \\
g^{(2)}_{3D} &=& 2\,\kappa\,g^{(1)}_{3D}\,, \\
g^{(2)}_{3E} &=& \kappa\,g^{(1)}_{3E}\,, \\
g^{(2)}_{3F} &=& 2\,\kappa\,g^{(1)}_{3F}\,.
\end{eqnarray}
\end{subequations}
The total two-loop Bethe-logarithm contribution is
\begin{equation}
g^{(2)}_{3} = g^{(2)}_{3A} + g^{(2)}_{3B} + g^{(2)}_{3C} +
              g^{(2)}_{3D} + g^{(2)}_{3E} + g^{(2)}_{3F}\,.
\label{71}
\end{equation}

The last part  of Eq.~(\ref{53}) $g^{(2)}_{4}$ involves the vacuum-polarization
correction. The contribution of the
diagrams with the closed fermion loop on the self-energy photon is
accounted for by the corresponding parts of the
electromagnetic form factors $F_1$, $F_2$ and $\eta$, $\xi$.
The two-loop vacuum polarization correction can be obtained
from the correction due to $F_1'^{(2)}(0)$ by the replacement
\begin{equation}
F_1'^{(2)}(0) \rightarrow v^{(2)} =
\biggl(\frac{\alpha}{\pi}\biggr)^2\,
\biggl(-\frac{82}{81}\times\frac{1}{4}\biggr)\label{64}\,.
\end{equation}
The corresponding contribution to the $g$ factor is
\begin{equation}
g^{(2)}_{4A} = -\biggl(\frac{\alpha}{\pi}\biggr)^2\,
\frac{82}{81}\, \langle 4\,\pi\,Z\,\alpha\,\delta^d(r)\rangle\,.
\label{65}
\end{equation}
The mixed self-energy and vacuum-polarization correction can be obtained
in a similar way by the replacement
\begin{equation}
F_2'^{(2)}(0) \rightarrow F_2^{(2)}(0)\,v^{(1)} =
F_2^{(2)}(0)\,\frac{\alpha}{\pi}\,\biggl(-\frac{1}{15}\biggr) \,.
\label{66}
\end{equation}
The corresponding contribution to the $g$ factor is
\begin{equation}
g^{(2)}_{4B} = \biggl(\frac{\alpha}{\pi}\biggr)^2\,
\frac{1}{15}\, \langle 4\,\pi\,Z\,\alpha\,\delta^d(r)\rangle \,.
\label{67}
\end{equation}
The total vacuum-polarization contribution
beyond the one accounted for by the form factors and $\eta$, $\xi$ is

\begin{equation}
g^{(2)}_{4} =g^{(2)}_{4A} +  g^{(2)}_{4B}\,.
\label{68}
\end{equation}

Finally, the complete two-loop correction to the bound-electron
$g$ factor is given by the sum of four parts in Eq.~(\ref{53}), which yields
\begin{widetext}
\begin{eqnarray}
g^{(2)} &=&
 \biggl(\frac{\alpha}{\pi}\biggr)^2\,\frac{(Z\,\alpha)^4}{n^3}\,\biggl\{
 \frac{28}{9}\,\ln[(Z\,\alpha)^{-2}]  +
   \frac{258917}{19440} - \frac{4}{9}\,\ln k_0 -
   \frac{8}{3}\,\ln k_3 + \frac{113}{810}\,{\pi }^2
 \nonumber \\ &&
 -  \frac{379}{90}\,{\pi }^2\,\ln 2
  + \frac{379}{60}\,\zeta(3)
   + \frac{1}{n}\left[
   -\frac{985}{1728}   - \frac{5}{144}\,{\pi }^2 +
      \frac{5}{24}\,{\pi }^2\,\ln 2 - \frac{5}{16}\,\zeta(3)\right]
  \biggr\}\,.
\label{72}
\end{eqnarray}
\end{widetext}
The numerical values for $\ln k_0$ and $\ln k_3$ for the first 7 $S$
states are
\begin{subequations}
\label{73}
\begin{align}
\label{lnk01S}
\ln k_0(1{S}) &= 2.984~128~556\,,\;\;
\ln k_3(1{S}) = 3.272~806~545\,,\\[1ex]
\label{lnk02S}
\ln k_0(2{S}) &= 2.811~769~893\,,\;\;
\ln k_3(2{S}) = 3.546~018~666\,,\\[1ex]
\label{lnk03S}
\ln k_0(3{S}) &= 2.767~663~612\,,\;\;
\ln k_3(3{S}) = 3.881~960~979\,,\\[1ex]
\label{lnk04S}
\ln k_0(4{S}) &= 2.749~811~840\,,\;\;
\ln k_3(4{S}) = 4.178~190~961\,,\\[1ex]
\label{lnk05S}
\ln k_0(5{S}) &= 2.740~823~727\,,\;\;
\ln k_3(5{S}) = 4.433~243~558\,,\\[1ex]
\label{lnk06S}
\ln k_0(6{S}) &= 2.735~664~206\,,\;\;
\ln k_3(6{S}) = 4.654~608~237\,,\\[1ex]
\label{lnk07S}
\ln k_0(7{S}) &= 2.732~429~129\,,\;\;
\ln k_3(7{S}) = 4.849~173~615\,,
\end{align}
\end{subequations}
The total numerical value of the nonlogarithmic term in Eq.~(\ref{72})
for the $1S$ state is $-16.436\,842$. All terms involving
the closed fermion loop contribute $-3.278\,177$ to this result, with the dominant
contribution originating from the two-loop vacuum-polarization
correction $g_{4A}^{(2)}$.

%
%
\begingroup
\squeezetable
\begin{table*}[htb]
\begin{center}
\begin{minipage}{16.0cm}
\caption{Individual contributions to the $1s$ bound-electron
$g$ factor. The abbreviations used are as follows: ``h.o.'' stands for
a higher-order contribution, ``SE'' -- for the self-energy
correction,
``VP-EL'' -- for the electric-loop vacuum-polarization correction,
``VP-ML'' -- for the magnetic-loop vacuum-polarization correction,
``TW'' indicates the results obtained in this work.
$\langle r^2\rangle^{1/2}$ is the root-mean-square nuclear
charge radius.
\label{tab:gfact} }
\begin{ruledtabular}
\begin{tabular}{ll...c}
&
& \multicolumn{1}{c}{$^{12}{\rm C}^{5+}$ }
& \multicolumn{1}{c}{$^{16}{\rm O}^{7+}$}
& \multicolumn{1}{c}{$^{40}{\rm Ca}^{19+}$}
& Ref. \\
\hline
$\langle r^2\rangle^{1/2}$[fm] &           & \multicolumn{1}{c}{2.4703\,(22)}  &  \multicolumn{1}{c}{2.7013\,(55)} & \multicolumn{1}{c}{3.4764\,(10)} & \cite{angeli:04} \\
\multicolumn{2}{l}{Dirac value (point nucleus)}&  1.998\, 721x\, 354\, 39\,(1)    & 1.997\, 726x\, 003\, 06\,(2)   &   1.985\, 723x\, 203\, 7\,(1)  & \\
Finite nuclear size  &                 &  0.000\, 000x\, 000\, 41         & 0.000\, 000x\, 001\, 55\,(1)   &   0.000\, 000x\, 113\, 0\,(1)  & \\
1-loop QED      & $     (Z\,\alpha)^0$ &  0.002\, 322x\, 819\, 47\,(1)    & 0.002\, 322x\, 819\, 47\,(1)   &   0.002\, 322x\, 819\, 5       &    \\
                & $     (Z\,\alpha)^2$ &  0.000\, 000x\, 742\, 16         & 0.000\, 001x\, 319\, 40        &   0.000\, 008x\, 246\, 2       & \cite{grotch:70:prl}\\
                & $     (Z\,\alpha)^4$ &  0.000\, 000x\, 093\, 42         & 0.000\, 000x\, 240\, 07        &   0.000\, 002x\, 510\, 6       & \cite{pachucki:04:prl} \\
                & h.o., SE             &  0.000\, 000x\, 008\, 28         & 0.000\, 000x\, 034\, 43\,(1)   &   0.000\, 003x\, 107\, 7\,(2)  & \cite{yerokhin:02:prl}+\cite{pachucki:04:prl} \\
                & h.o., VP-EL          &  0.000\, 000x\, 000\, 56         & 0.000\, 000x\, 002\, 24        &   0.000\, 000x\, 172\, 7       & \cite{beier:00:rep} \\
                & h.o., VP-ML          &  0.000\, 000x\, 000\, 04         & 0.000\, 000x\, 000\, 16        &   0.000\, 000x\, 014\, 6       & \cite{lee:05:pra} \\
$\ge$2-loop QED &$     (Z\,\alpha)^0$  & -0.000\, 003x\, 515\, 10         &-0.000\, 003x\, 515\, 10        &  -0.000\, 003x\, 515\, 1       & \cite{mohr:05:rmp}\\
                &$     (Z\,\alpha)^2$  & -0.000\, 000x\, 001\, 12         &-0.000\, 000x\, 002\, 00        &  -0.000\, 000x\, 012\, 5       & \cite{grotch:70:prl}\\
                &$     (Z\,\alpha)^4$  &  0.000\, 000x\, 000\, 06         & 0.000\, 000x\, 000\, 08        &  -0.000\, 000x\, 010\, 9       & TW \\
                & h.o.                 &  0.000\, 000x\, 000\, 00\,(3)    & 0.000\, 000x\, 000\, 00\,(11)  &   0.000\, 000x\, 000\, 0\,(100)&   \\
Recoil          &$ m/M$                &  0.000\, 000x\, 087\, 70         & 0.000\, 000x\, 117\, 07        &   0.000\, 000x\, 297\, 3       & \cite{shabaev:02:prl} \\
                & h.o.                 & -0.000\, 000x\, 000\, 08         &-0.000\, 000x\, 000\, 10        &  -0.000\, 000x\, 000\, 3       & \cite{martynenko:01:jetp}  \\
Total           &                      &  2.001\, 041x\, 590\, 18\,(3)    & 2.000\, 047x\, 020\, 32\,(11)  &   1.988\, 056x\, 946\, 6\,(100)  \\
\end{tabular}
\end{ruledtabular}
\end{minipage}
\end{center}
\end{table*}
\endgroup

\section{Results and discussion}

In Table~\ref{tab:gfact}, we collect all contributions available for the 
$1S$ bound-electron $g$ factor in three specific hydrogenlike ions which are 
important from an experimental point of view. For two of them, 
carbon and oxygen, accurate experimental results are presently available 
\cite{haeffner:00:prl,verdu:04}, whereas the experiment on calcium is 
planned for the future \cite{quint:04:priv}.

The errors of the point-nucleus Dirac value and of the free part of the 
one-loop QED correction indicated in the table originate from the 
uncertainty of the fine-structure constant, $\alpha^{-1} = 
137.035\,999\,11(46)$ \cite{mohr:05:rmp}. The finite-nuclear-size 
correction was re-evaluated in this work using the most recent values for 
the root-mean-square (rms) nuclear radii~\cite{angeli:04}. The error 
ascribed to this correction originates both from the uncertainty of the 
rms radius and from the estimated model dependence for the nuclear-charge 
distribution.

The one-loop QED correction up to the order of $(Z\,\alpha)^4$ is given by the sum of 
the self-energy part [Eq.~(\ref{52})] and the vacuum-polarization 
part~\cite{karshenboim:00:pla},
\begin{equation}
\label{74}
g^{(1)}_{\rm VP} = \frac{\alpha}{\pi}\, (Z\,\alpha)^4\,
    \left( -\frac{16}{15} \right)\,.
\end{equation}
The higher-order one-loop self-energy correction was inferred
from the results of the all-order numerical calculation
\cite{yerokhin:02:prl,yerokhin:04:pra}. For carbon and oxygen, the results
presented in the table
were obtained in Ref.~\cite{pachucki:04:prl} by an extrapolation
of the numerical results \cite{yerokhin:02:prl} for $Z>8$, after subtracting
the known terms of the $Z \alpha$ expansion.
The one-loop vacuum-polarization correction consists of two parts, the
electric-loop contribution that is due to the vacuum-polarization
insertion into the electron line and the magnetic-loop contribution,
which corresponds to the insertion of the vacuum-polarization loop
into the interaction with the external magnetic field. The values for the
higher-order electric-loop contribution presented in the table were
inferred from the all-order numerical results of Ref.~\cite{beier:00:rep},
whereas the magnetic-loop contribution was taken from the recent evaluation
\cite{lee:05:pra}.

The $(Z\,\alpha)^0$ and $(Z\,\alpha)^2$ parts of the two- and more-loop QED 
correction comprise the two-, three-, and four-loop contributions to the 
free-electron $g$ factor, multiplied by a kinematic factor of the electron
\cite{grotch:70:prl}. The $(Z\,\alpha)^4$ part of the two-loop QED 
contribution was derived in the present work. The uncertainty due to 
higher-order two-loop contributions was estimated as
\begin{equation}
g^{(2)}_{\rm h.o.} = 2\, g^{(1)}_{\rm h.o.}\,
     \frac{g^{(2)}[(Z\,\alpha)^2]}{g^{(1)}[(Z\,\alpha)^2]}\,,
\end{equation}
where $g^{(n)}_{\rm h.o.}$ is the $n$-loop higher-order QED contribution
and $g^{(n)}[(Z\,\alpha)^2]$ is the $n$-loop $(Z\,\alpha)^2$ QED contribution.

The nuclear recoil correction to first order in the mass ratio $m/M$
but to all order in $Z\,\alpha$ was calculated in 
Refs.~\cite{shabaev:01:rec,shabaev:02:prl}. The leading recoil 
corrections to order $(m/M)^2$ and $\alpha\,m/M$ were derived in 
Refs.~\cite{grotch:70:pra,faustov:70} for a nuclear spin $I=1/2$ and 
recently generalized for an arbitrary nuclear spin in 
Ref.~\cite{martynenko:01:jetp}.

Based on the data presented in Table~\ref{tab:gfact}, we conclude that 
our evaluation of the one- and two-loop QED corrections to order 
$(Z\,\alpha)^4$ improves the accuracy of the theoretical prediction for carbon 
by an order of magnitude, as compared to the previous compilation 
\cite{yerokhin:02:prl}. The resulting QED contribution to order $(Z\,\alpha)^4$ 
turns out to be rather small for carbon and oxygen, as a result 
of a cancellation between the logarithmic and the 
nonlogarithmic parts of this correction [see Eq.~(\ref{72})]. 
For calcium, to the contrary, the numerical contribution 
of the two-loop $(Z\,\alpha)^4$ correction is large and of the same order as 
the $(Z\,\alpha)^2$ correction. This indicates that the perturbative 
$Z\,\alpha$-expansion approach is no longer effective in this region of $Z$,
and a direct all-order numerical evaluation would be 
highly desirable. 

It is remarkable that among different sources of the theoretical 
uncertainty for calcium, the error due to the higher-order two-loop QED 
correction is by far the dominant one. This means that, if the 
prospective experimental investigation of the bound-electron $g$ factor 
in calcium is performed on the same level of accuracy as for carbon, 
namely $10^{-9}$, a comparison of the theoretical and experimental 
results would allow one to identify the contribution of the 
non-perturbative (in $Z\,\alpha$) two-loop QED effects with a 10\% accuracy.

The comparison of the theoretical and experimental results for the $1S$ 
bound-electron $g$ factor in carbon and oxygen yields the presently most 
accurate method for determination of the electron 
mass~\cite{mohr:05:rmp}. Based on the theoretical $g$ factor values presented 
in Table~\ref{tab:gfact}, we obtain the following values for the electron 
mass derived from the experiments on carbon \cite{haeffner:00:prl} and 
oxygen \cite{verdu:04} (in atomic mass units):
\begin{eqnarray}
m(^{12}{\rm C}^{5+}) &=& 0.000\, 548\, 579\, 909\, 32\, (29)\,,
    \\
m(^{16}{\rm O}^{7+}) &=& 0.000\, 548\, 579\, 909\, 60\, (41)\,.
\end{eqnarray}
The uncertainty of these results originates from the experimental
value for the ratio of the electronic Larmor precession frequency
and the cyclotron frequency of the ion in the trap; the
uncertainty due to the theoretical prediction is more than by an
order of magnitude smaller and thus negligible.

\section*{Acknowledgments}

Valuable discussions with W.~Quint are gratefully acknowledged.
This work was supported by EU grant No.~HPRI-CT-2001-50034 and by RFBR
grant No.~04-02-17574. C.A. acknowledges the support by the 
Natural Sciences and Engineering Research Canada.  
V.A.Y. acknowledges support by the
foundation ``Dynasty''. U.D.J. acknowledges support from the 
Deutsche Forschungsgemeinschaft via the Heisenberg program.

\appendix
\section{Electromagnetic form factors}
\label{appa}

We consider the form factors defined by
\begin{equation}
\gamma_\mu \to \Gamma_\mu = 
F_1(q^2)\gamma_\mu + {i\over 2m} F_2(q^2) \left( {i\over 2} \right)[\qsla, \gamma_\mu]\,,
\label{A1}
\end{equation}
where $q$ is the outgoing photon momentum. The form factors are expanded in
$\alpha$ up to second order,
\begin{eqnarray}
F_1(q^2) &=& 1 +  F_1^{(1)}(q^2)+ F_1^{(2)}(q^2)\,,
\nonumber\\
F_2(q^2) &=&  F_2^{(1)}(q^2)+ F_2^{(2)}(q^2),
\end{eqnarray}
where the superscript corresponds to the loop order,
i.e.~to the power of $\alpha$.
The results for the form factors expanded into powers of $q^2$ up to
$q^4$ read (in $D=4-2\varepsilon$):
\begin{widetext}
\begin{subequations}
\label{ff}
\begin{eqnarray}
F_1^{(1)}(q^2) &=& \frac{\alpha}{\pi}\,\biggl[q^2 \left( - {1 \over 8} 
- {1 \over 6\varepsilon} - {1 \over 2} \varepsilon \right)
+ q^4 \left( - {11 \over 240} 
- {1 \over 40\varepsilon} 
- {5 \over 48} \varepsilon \right)\biggr]\,,\\
F_2^{(1)}(q^2) &=&  \frac{\alpha}{\pi}\,\biggl[
{1 \over 2} + 2 \varepsilon
+ q^2 \left( {1 \over 12} + {5 \over 12} \varepsilon  \right)
+ q^4 \left( {1 \over 60} + {11 \over 120} \varepsilon  \right)\biggr]\,,\\
F_1^{(2)}(q^2) &=& \Bigl(\frac{\alpha}{\pi}\Bigr)^2\,\biggl\{
q^2 \left[ 
\left( -\frac{1099}{1296} + \frac{77}{144} \zeta(2)\right)_{\rm VP}
- {47 \over 576} 
+ 3 \, \zeta(2) \, \ln 2 
- {175 \over 144} \zeta(2)
- {3 \over 4}\zeta(3) 
\right]
\\
&& + q^4 \left[  \left( -\frac{491}{1440}  + \frac{5}{24} \zeta(2)\right)_{\rm VP}
+ {1721 \over 12960} 
+ {1 \over 72 \, \varepsilon^2}
+ {1 \over 48 \, \varepsilon}
+ {11 \over 10} \, \zeta(2) \, \ln 2 
- {14731 \over 28800} \, \zeta(2) 
- {11 \over 40} \zeta(3) \right]\biggr\} \,,
\nonumber\\
F_2^{(2)}(q^2) &=&  \Bigl(\frac{\alpha}{\pi}\Bigr)^2\,\biggl\{
\left( \frac{119}{36} - 2 \zeta(2)\right)_{\rm VP}-{31 \over 16} 
- 3 \, \zeta(2) \, \ln 2 
+ {5 \over 2} \, \zeta(2) 
+ {3 \over 4} \, \zeta(3)
\nonumber\\ 
&& + q^2 \left[ \left( \frac{311}{216}  - \frac{7}{8} \zeta(2)\right)_{\rm VP}
-{77 \over 80} 
- {1 \over 12\,\varepsilon} 
- {23 \over 10} \, \zeta(2) \, \ln 2
+ {61 \over 40} \, \zeta(2) 
+ {23 \over 40} \, \zeta(3) 
\right]
\nonumber\\
&& + q^4 \left[ \left( \frac{533}{1080} - \frac{3}{10} \zeta(2)\right)_{\rm VP}
-{1637 \over 5040} 
- {19 \over 720\,\varepsilon}
- {15 \over 14} \, \zeta(2) \, \ln 2 
+ {689 \over 1050} \, \zeta(2) 
+ {15 \over 56} \, \zeta(3) \right]\biggr\}\,.
\end{eqnarray}
\end{subequations}
\end{widetext}
The subscript VP denotes the contribution to the two-loop
form factors which involves a closed fermion loop.

\section{The low-energy limit of the scattering amplitude}
\label{appb}

\begin{figure}[htb]
\includegraphics[width=1.0\linewidth]{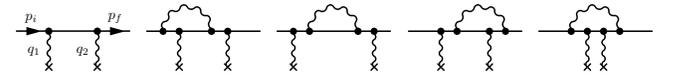}
\caption{\label{fig1}
Feynman diagrams representing the scattering amplitude of a free
electron on both the Coulomb and the magnetic field, at the tree and
the one-loop level. }
\end{figure}

In this section we describe the  evaluation of 
the low-energy limit of the spin-dependent part
of the scattering amplitude that gives rise to the effective
Hamiltonian (\ref{17}). The scattering amplitude under 
consideration is schematically depicted in
Fig.~\ref{fig1}, where the leftmost graph is the ``tree''  diagram
and the remaining graphs represent the tree diagram ``dressed'' by
a self-energy photon. The two-loop diagrams are not shown explicitly;
they can be obtained from the one-photon ones in a standard way.
Each graph contains two interactions with the external field, one of
which is the interaction with the homogeneous magnetic field
(a $\gamma^i$ vertex) and the other is the interaction with the
Coulomb field of the nucleus (a $\gamma^0$ vertex).
From the one- and two-loop scattering amplitudes we additionally
subtract a tree amplitude with the vertices modified by the electromagnetic
form factors $F_1$ and $F_2$. This procedure removes the part that is
already accounted for by the Hamiltonian (\ref{16}) and  leads to
a simple polynomial expression for the resulting amplitude.

In order to extract the spin dependent part of the scattering amplitude,
we construct the projection operator. Let us first consider 
a general non-relativistic operator $Q$,
\begin{equation}
Q = Q^0 + Q^i\,\sigma^i\,.
\end{equation}
The spin-dependent part of $Q$ can be retrieved by the following projection
operator:
\begin{equation}
Q^i = \frac{1}{2}\,{\rm Tr}[Q\,\sigma^i]\,.
\end{equation}
In $d$ dimensions, the nonrelativistic expansion of the Hamiltonian
involves $\sigma^{ij} = [\sigma^i\,,\sigma^j]/(2\,{\rm i})$.
The extension of the spin-projection operator to
an arbitrary number of dimensions is
\begin{equation}
Q^{ij} = \frac{1}{4}\,{\rm Tr}[Q\,\sigma^{ij}]\,,
\end{equation}
with $Q = Q^{ij}\,\sigma^{ij}$. 
We assume here the following properties of the trace to hold:
\begin{eqnarray}
{\rm Tr}[\sigma^{ij}] &=& 0\,, \nonumber \\
{\rm Tr}[{\rm I}] &=& 2\,, \nonumber \\
{\rm Tr}[\sigma^{ij}\,\sigma^{kl}] &=& 2\,(\delta^{ik}\,\delta^{jl}-\delta^{jk}\,\delta^{il})\,.
\end{eqnarray}

We now consider the operator $Q$ sandwiched between the positive-energy
solutions of the free Dirac equation normalized by $\bar u \,u=1$.
The following identity holds,
\begin{equation}
\bar u(p_{\rm f},s_{\rm f})\,Q\,u(p_{\rm i},s_{\rm i}) 
= {\rm Tr}[Q\,u(p_{\rm i},s_{\rm i})\,\bar u(p_{\rm f},s_{\rm f})]\,.
\end{equation}
Since our aim is to calculate the low-energy limit of the amplitude only, we can use
an approximate form for $u(p,s)$,
\begin{eqnarray}
u(p,s) \approx \left(
\begin{array}{l}
\phi(s)\\[2ex]
\frac{1}{2}\,\vec{\sigma}\cdot\vec{p}\,\phi(s)
\end{array}
\right)\,,
\end{eqnarray}
where $\phi$ is a nonrelativistic spinor.
Using a replacement that extracts the spin dependence
\begin{equation}
\phi(s_{\rm i})\,\phi^+(s_{\rm f}) \rightarrow \frac{\sigma^{ij}}{4}\,,
\end{equation}
the projection operator becomes (in units $m=1$)
\begin{align}
& u(p_{\rm i},s_{\rm i})\,\bar u(p_{\rm f},s_{\rm f}) \rightarrow \frac{1}{4}\left(
\begin{array}{lr}
\sigma^{ij},&-\frac{1}{2}\,\sigma^{ij}\,\vec{\sigma}\cdot\vec{p}_{\rm f}\\
\frac{1}{2}\,\vec{\sigma}\cdot\vec{p}_{\rm i}\,\sigma^{ij},&
-\frac{1}{4}\,\vec{\sigma}\cdot\vec{p}_{\rm i}\,\,\sigma^{ij}\,
\vec{\sigma}\cdot\vec{p}_{\rm f}
\end{array}\right)
\nonumber\\
& \approx \frac{1}{16}\,(\not\!p_{\rm i}+1)\,\Sigma^{ij}\,(\not\!p_{\rm f}+1)\,.
\end{align}
Therefore,
\begin{equation}
Q^{ij}  = \frac{1}{16}\,{\rm Tr}\left[
(\not\!p_{\rm f}+1)\,Q\,(\not\!p_{\rm i}+1)\,\Sigma^{ij}
\right]\,.
\end{equation}

We now turn to the scattering amplitude of the free electron on the Coulomb and
magnetic fields. The spin-dependent part of this amplitude
is written as
\begin{equation}
Q = Q^{\mu\nu\rho}\,e\,A_0(q_1)\,e\,A_\mu(q_2)\,\sigma_{\nu\rho}\,,
\label{B10}
\end{equation}
where $q_1$ and $q_2$ denote the exchange momenta.
The amplitude corresponding to the tree diagram in Fig.~\ref{fig1} is
given by
\begin{align}
& Q^{\mu\nu\rho}_0 =
\frac{1}{16}\,{\rm Tr}
\left[(\not\!p_{\rm f}+1)\,\gamma^0\,\frac{1}{\not\!p_{\rm i}+\not\!q_2-1}\,
\gamma^\mu\,(\not\!p_{\rm i}+1)\,\Sigma^{\nu\rho} \right.
\nonumber\\ 
& \quad \left. +(\not\!p_{\rm f}+1)\,\gamma^\mu\,\frac{1}{\not\!p_{\rm i}+\not\!q_1-1}\,
\gamma^0\,(\not\!p_{\rm i}+1)\,\Sigma^{\nu\rho}\right]\,,
\end{align}
where the momenta $p_{\rm i}$, $p_{\rm f}$ are on the mass shell, and
the exchange momenta are spatial, $q_1^0 = q_2^0 = 0$.

As an example of one-photon contributions, we give an expression for
the rightmost diagram in Fig.~\ref{fig1},
%
\begin{align}
& Q^{\mu\nu\rho}_1 = -i\,e^2\int\frac{d^D k}{(2\,\pi)^D}\,\frac{1}{k^2}\,
\frac{1}{16}\,{\rm Tr} \Biggl[(\not\!p_{\rm f}+1)\, \gamma^\sigma
\nonumber \\ 
& \times \frac{1}{\not\!p_{\rm f}-\not\!k-1}\,
\,\gamma^0\, \frac{1}{\not\!p_{\rm i}+\not\!q_2-\not\!k-1} \,\gamma^\mu\,
\frac{1}{\not\!p_{\rm i}-\not\!k-1}\,
\nonumber \\ 
& \times \gamma_\sigma
(\not\!p_{\rm i}+1)\,\Sigma^{\nu\rho}\Biggr] +{\rm symmetrization}\,.
\end{align}
%
The other one- and two-loop contributions are obtained in the analogous way.
From the resulting amplitude we subtract
the tree amplitude $Q^{\mu\nu\rho}_F$ with 
vertices $\gamma^\alpha$ replaced by $\Gamma^\alpha$,
\begin{align}
& Q^{\mu\nu\rho}_{F} =
\nonumber \\ 
& \frac{1}{16}\,
{\rm Tr}\left[(\not\!p_{\rm f}+1)\,\Gamma^0(q_1)\,\frac{1}{\not\!p_{\rm i}+\not\!q_2-1}\,
\Gamma^\mu(q_2)\,(\not\!p_{\rm i}+1)\,\Sigma^{\nu\rho}
\right.
\nonumber\\ 
& \left.
+(\not\!p_{\rm f}+1)\,\Gamma^\mu(q_2)\,\frac{1}{\not\!p_{\rm i}+\not\!q_1-1}\,
\Gamma^0(q_1)\,(\not\!p_{\rm i}+1)\,\Sigma^{\nu\rho}\right] \,,
\end{align}
where $\Gamma^\alpha$ is defined in Eq. (\ref{A1}).
The final expression for the total amplitude  $Q^{\mu\nu\rho}$ is obtained by
the expansion in small momenta $\vec{p}_{\rm i}\,,\vec{p}_{\rm f}$ and the
subsequent integration over the loop momenta.
The result for $Q^{\mu\nu\rho}$ can be written in the form
\begin{eqnarray} 
Q^{\mu\nu\rho} =
\frac{1}{2}\left[ \eta\; {\cal F}^{\mu\nu\rho} 
+ \xi\; {\cal G}^{\mu\nu\rho}\right],
\end{eqnarray} 
where the functions ${\cal F}^{\mu\nu\rho}$ and 
${\cal G}^{\mu\nu\rho}$ are orthogonal to 
$q_2^\mu$ (due to the gauge invariance)
and antisymmetric in $\nu,\rho$. Their explicit expressions are
\begin{eqnarray} 
{\cal F}^{\mu\nu\rho} &= &
q_1^\mu \left(q_1^\rho q_2^\nu - q_1^\nu q_2^\rho\right)
+ q_1\cdot q_2 \left(g^{\mu \rho} q_1^\nu-g^{\mu \nu} q_1^\rho\right)\,,
\nonumber \\
{\cal G}^{\mu\nu\rho} &= &
q_1^2 \left(g^{\mu \rho} q_2^\nu-g^{\mu \nu} q_2^\rho\right)\,.
\end{eqnarray} 
The results for the  coefficient functions $\eta$ and $\xi$ read
\begin{align} 
\label{eta}
\eta =& -{\alpha\over 4\pi} {  2 \over 3 \varepsilon}
+ \left({\alpha\over 4\pi}\right)^2  \left[
\left( {2528 \over 81} - {169 \over 54}\pi^2\right)_{\rm VP}
\right.
\nonumber\\ 
& \left. 
-{283 \over 10}  + {169 \over 120}\,\pi^2 - {4 \over 15}\,\pi^2 \ln 2 +
{2  \over 5}\,\zeta(3) - {16 \over 3 \varepsilon}\right], \\
\label{xi}
\xi =& {\alpha\over 4\pi} \left( 1 +{ 2 \over 3 \varepsilon} \right)
+ \left({\alpha\over 4\pi}\right)^2 
\left[ \left({ 2674 \over 81}-{91 \over 27}\pi^2\right)_{\rm VP} \right.
\nonumber \\ 
& \left.  -{152 \over 15} + {319 \over 45}\,\pi^2 -{ 68  \over 5}\,\pi^2\ln 2
+ {102  \over 5}\,\zeta(3)  + {4 \over 3  \varepsilon}
\right]\,,
\end{align}
where the subscript VP denotes the contribution involving a closed fermion loop.
The effective local operator $Q$ in Eq.~(\ref{B10}) becomes
\begin{align}
Q &=& \frac{1}{2}\left[
\eta\; {\cal F}^{\mu\nu\rho} + \xi\; {\cal G}^{\mu\nu\rho}\right]
\,e\,A_0(q_1)\,e\,A_\mu(q_2)\,\sigma_{\nu\rho}\nonumber \\
&\rightarrow& \frac{e^2}{2\,m}\,\left[2\,\sigma^{ij}\,B^{ik}\,\nabla^j E^k\,\eta
  + \sigma^{ij} B^{ij}\,\nabla^k E^k\,\xi\right]\,,
\end{align}
which corresponds to the effective Hamiltonian in Eq. (\ref{17}).


\begin{thebibliography}{10}

\bibitem{hughes:99}
V.~W. Hughes and T.~Kinoshita,
\newblock Rev. Mod. Phys. {\bf 71}, 133  (1999).

\bibitem{mohr:05:rmp}
P.~J. Mohr and B.~N. Taylor,
\newblock Rev. Mod. Phys. {\bf 77}, 1  (2005).

\bibitem{beier:02:prl}
T.~Beier, H.~H\"affner, N.~Hermanspahn, S.~G. Karshenboim, H.-J. Kluge,
  W.~Quint, S.~Stahl, J.~Verd\'u, and G.~Werth,
\newblock Phys. Rev. Lett. {\bf 88}, 011603 (2002).

\bibitem{haeffner:00:prl}
H.~H\"{a}ffner, T.~Beier, N.~Hermanspahn, H.-J. Kluge, W.~Quint, S.~Stahl,
  J.~Verd\'{u}, and G.~Werth,
\newblock Phys. Rev. Lett. {\bf 85}, 5308  (2000).

\bibitem{verdu:04}
J.~Verd\'u, S.~Djekic, S.~Stahl, T.~Valenzuela, M.~Vogel, G.~Werth, T.~Beier,
  H.-J. Kluge, and W.~Quint,
\newblock Phys. Rev. Lett. {\bf 92}, 093002 (2004).

\bibitem{breit:28}
G.~Breit,
\newblock Nature (London) {\bf 122}, 649 (1928).

\bibitem{blundell:97:pra}
S.~A. Blundell, K.~T. Cheng, and J.~Sapirstein,
\newblock Phys. Rev. A {\bf 55}, 1857 (1997).

\bibitem{persson:97:g}
H.~Persson, S.~Salomonson, P.~Sunnergren, and I.~Lindgren,
\newblock Phys. Rev. A {\bf 56}, R2499  (1997).

\bibitem{beier:00:pra}
T.~Beier, I.~Lindgren, H.~Persson, S.~Salomonson, P.~Sunnergren,
  H.~H\"{a}ffner, and N.~Hermanspahn,
\newblock Phys. Rev. A {\bf 62}, 032510 (2000).

\bibitem{yerokhin:02:prl}
V.~A. Yerokhin, P.~Indelicato, and V.~M. Shabaev,
\newblock Phys. Rev. Lett. {\bf 89}, 043001 (2002).

\bibitem{yerokhin:04:pra}
V.~A. Yerokhin, P.~Indelicato, and V.~M. Shabaev,
\newblock Phys. Rev. A {\bf 69}, 052503 (2004).

\bibitem{karshenboim:02:plb}
S.~G. Karshenboim and A.~I. Milstein,
\newblock Phys. Lett. B {\bf 549}, 321 (2002).

\bibitem{lee:05:pra}
R.~N. Lee, A.~I. Milstein, I.~S. Terekhov, and S.~G. Karshenboim,
\newblock Phys. Rev. A {\bf 71}, 052501 (2005).

\bibitem{shabaev:01:rec}
V.~M. Shabaev,
\newblock Phys. Rev. A {\bf 64}, 052104 (2001).

\bibitem{martynenko:01:jetp}
A.~P. Martynenko and R.~N. Faustov,
\newblock Zh. Eksp. Teor. Fiz. {\bf 120}, 539  (2001)
\newblock [JETP {\bf 93}, 471 (2001)].

\bibitem{shabaev:02:prl}
V.~M. Shabaev and V.~A. Yerokhin,
\newblock Phys. Rev. Lett. {\bf 88}, 091801 (2002).

\bibitem{nefiodov:02:prl}
A.~V. Nefiodov, G.~Plunien, and G.~Soff,
\newblock Phys. Rev. Lett. {\bf 89}, 081802 (2002).

\bibitem{glazov:04:pra}
D.~A. Glazov, V.~M. Shabaev, I.~I. Tupitsyn, A.~V. Volotka, V.~A. Yerokhin,
  G.~Plunien, and G.~Soff,
\newblock Phys. Rev. A {\bf 70}, 062104 (2004).

\bibitem{moskovkin:04}
D.~L. Moskovkin, N.~S. Oreshkina, V.~M. Shabaev, T.~Beier, G.~Plunien,
  W.~Quint, and G.~Soff,
\newblock Phys. Rev. A {\bf 70}, 032105 (2004).

\bibitem{pachucki:04:prl}
K.~Pachucki, U.~D. Jentschura, and V.~A. Yerokhin,
\newblock Phys. Rev. Lett. {\bf 93}, 150401 (2004);
\newblock erratum Phys. Rev. Lett. {\bf 94}, 229902(E) (2005).

\bibitem{czarnecki:99:prl}
A.~Czarnecki, K.~Melnikov, and A.~Yelkhovsky,
\newblock Phys. Rev. Lett. {\bf 82}, 311  (1999).

\bibitem{korobov:01:prl}
V.~Korobov and A.~Yelkhovsky,
\newblock Phys. Rev. Lett. {\bf 87}, 193003 (2001).

\bibitem{itzykson:80}
C.~Itzykson and J.~B. Zuber,
\newblock {\em Quantum Field Theory},
\newblock Mc{G}raw-Hill, New York, 1980.

\bibitem{pachucki:04:lwqed}
K.~Pachucki,
\newblock Phys. Rev. A {\bf 69}, 052502 (2004).

\bibitem{shabaev:91:jpb}
V.~M. Shabaev,
\newblock J. Phys. B {\bf 24}, 4479  (1991).

\bibitem{shabaev:03:PSAS}
V.~M. Shabaev,
\newblock in {\em Precision Physics of Simple Atomic Systems}, ed. by S.~G.
  Karshenboim and V.~B. Smirnov, Springer, Berlin, 2003, p. 97.

\bibitem{quint:04:priv}
W.~Quint,
\newblock private communication (2004).

\bibitem{angeli:04}
I.~Angeli,
\newblock At. Data Nucl. Data Tables {\bf 87}, 185  (2004).

\bibitem{karshenboim:00:pla}
S.~G. Karshenboim,
\newblock Phys. Lett. A {\bf 266}, 380  (2000).

\bibitem{beier:00:rep}
T.~Beier,
\newblock Physics Reports {\bf 339}, 79  (2000).

\bibitem{grotch:70:prl}
H.~Grotch,
\newblock Phys. Rev. Lett. {\bf 24}, 39  (1970).

\bibitem{grotch:70:pra}
H.~Grotch,
\newblock Phys. Rev. A {\bf 2}, 1605  (1970).

\bibitem{faustov:70}
R.~Faustov,
\newblock Phys. Lett. B {\bf 33}, 422 (1970).

\end{thebibliography}

\end{document}